# Mechanical Modeling of Innovative Metamaterials Alternating Pentamode Lattices and Confinement Plates

F. Fraternali[1], A. Amendola[1]

[1]Department of Civil Engineering, University of Salerno, Via Giovanni Paolo II 132, 84084 Fisciano (SA), Italy. f.fraternali@unisa.it (F. Fraternali), adaamendola1@unisa.it (A. Amendola).

## Abstract

*This study examines the mechanical behavior of a novel class of mechanical metamaterials alternating pentamode lattices and stiffening plates. The unit cell of such lattices consists of a sub-lattice of the face cubic-centered unit cell typically analyzed in the current literature on pentamode materials. The studied systems exhibit only three soft deformation modes in the infinitesimal stretch-dominated regime, as opposed to the five zero-energy modes of unconfined pentamode lattices. We develop analytical formulae for the vertical and bending stiffness properties and study the dependence of such quantities on the main design parameters: the lattice constant, the solid volume fraction, the cross-section area of the rods, and the layer thickness. A noteworthy result is that the effective compression modulus of the analyzed structures is equal to two thirds of the Young modulus of the stiffest isotropic elastic networks currently available in the literature, being accompanied by zero-rigidity against infinitesimal shear and twisting mechanisms. The use of the proposed metamaterials as novel seismic-isolation devices and impact-protection equipment is discussed by drawing comparisons with the response of alternative devices already available or under development.*

## 1. Introduction

Pentamode lattices are mechanical metamaterials that exhibit the minimal coordination number required to achieve a fully positive definite elasticity tensor in three dimensions [1]-[5]. Their 2D counterparts are honeycomb lattices [4], or actuated bimode metamaterials [5]. Pentamode lattice materials are characterized by an elementary unit cell showing four rods converging at a point, and are known to exhibit five zero-energy modes of deformation [1], [4]. Their use as stop-band materials for shear waves and elasto-mechanical cloaks forms the subject of active ongoing research in several branches of mechanics and physics [6]-[9].

Recent studies have shown a novel feature of confined pentamode lattices, which consists of the capacity to carry unidirectional compressive loads with sufficiently high stiffness, while showing markedly low stiffness against shear loads [10]-[13]. While many cell unconfined pentamode lattices feature zero Young modulus in the stretch-dominated limit [1][4], other research [10]-[13] has shown that single- and multi-layer structures formed by the alternating pentamode lattices and stiffening plates are able to oppose a noticeable degree of rigidity to unidirectional compression loads in the bending-dominated regime, due to the confinement effect provided by the stiffening plates. Such a feature is essential when developing mechanical metamaterials that need to carry significantly large loads perpendicular to their outer surface while exhibiting low (theoretically zero) rigidity against transverse shear forces [10]. The research presented in [10]-[13] considers lattices formed by the repetition in the 3D space of a face-centered-cubic (fcc) unit cell composed of four primitive pentamode cells (*fcc lattices*). An inherent limitation of such systems is that their compression rigidity stems from the bending rigidity of nodes and rods, which completely vanishes in the stretch-dominated limit.

The present study analyzes pentamode lattices whose unit cell consists of a suitable sub-lattice of the fcc cell, being formed by only two primitive cells (sfcc lattices, cf. Sect. 2). Considering the infinitesimal incremental motions from the reference configuration of an elementary sfcc module (Sect. 3), we show that the examined systems, when equipped with perfectly hinged connections, feature only three zero-energy modes (Sect. 4), and exhibit positive elastic rigidity against both vertical and bending loads (Sect. 5). We conclude that single- and multi-layer structures alternating sfcc pentamode lattices and confinement plates are able to carry vertical and bending loads also in the presence of zero bending rigidity of nodes and rods, as opposed to confined and unconfined fcc systems. Sects. 5.1-5.3 provide analytical formulae for the vertical and bending stiffness properties of the analyzed metamaterials, and study the dependence of such quantities on the main design parameters, which include the lattice constant, the solid volume fraction, the cross-section area of the rods, and the layer thickness. It is worth noting, in particular, that in the linear elastic regime, perfectly hinged sfcc lattices exhibit an effective compression modulus equal to $2/3$ of the Young modulus of the stiffest elastic networks analyzed in Ref. [14]. We make some comparisons between the mechanical response of the pentamode materials analyzed in the present work and that of rubber bearings formed by elastomeric layers confined between stiffening plates. The finite element results given in Sect. 6 allow us to validate the analytic predictions of the stiffness coefficients of the sfcc structures presented in Sect. 5. Sect. 7 summarizes the key mechanical features of sfcc systems and suggests future research lines for the design and testing of physical models of such novel metamaterials, which show promise for the next generation of impact-protector equipment and base-isolation devices.

## 2. Layered sfcc pentamode lattices

Let us consider laminated structures composed of layers of pentamode lattices confined between stiffening plates. The extended face-centered-cubic (fcc) unit cell of a pentamode lattice, which is formed by four primitive unit cells comprising four rods meeting at a point, is shown in Figure 1a. The present study examines pentamode lattices obtained by repeating in the 3D space the sub-lattice of the fcc unit cell shown in Figure 1b, which is formed by two primitive cells. We name the unit cell *sfcc*, and use the term *sfcc lattices* for the structures obtained by repeating such a cell in the horizontal plane. Laminated sfcc structures are built by alternating in the vertical direction sfcc lattices and stiffening plates, as shown in Figure 1c. We hereafter examine pentamode lattices that are endowed with hinged connections and exhibit a pure stretching response (no bending deformation [4][10], Sects. 2-5). Such connections may, for instance, consist of the hollow ball joints commonly used in structural space grids [15]. For the sake of simplicity, we assume that both the layers of pentamode lattices and the stiffening plates of the examined systems exhibit uniform properties across the layered structure.

By referring the geometry of a layered sfcc structure to an $x, y, z$ Cartesian frame aligned with unit cell edges, such that the $x, y$ axes lie in the horizontal plane (Figure 1b), we let $L_x$, $L_y$ denote the edge lengths of the stiffening plates, and let $n_x, n_y$ and $n_z$ denote the number of unit cells placed along the $x, y$ and $z$ axes in the generic layer, respectively. In addition, we denote the height of the generic pentamode layer by $H_i$, and set $H = n_z H_i$ (total height of the pentamode layers). The total height of the overall laminated structure, which includes the thicknesses of the stiffening plates, is denoted by $\bar{H}$. We assume that the rods forming the pentamode lattices have constant cross-section with cross-section area $s$, and are formed by a homogeneous and linearly elastic material with Young modulus $E_0$. Concerning the stiffening plates, we instead assume that such elements behave as 2D rigid bodies during an arbitrary deformation of the structure, both in-plane and out-of-plane.

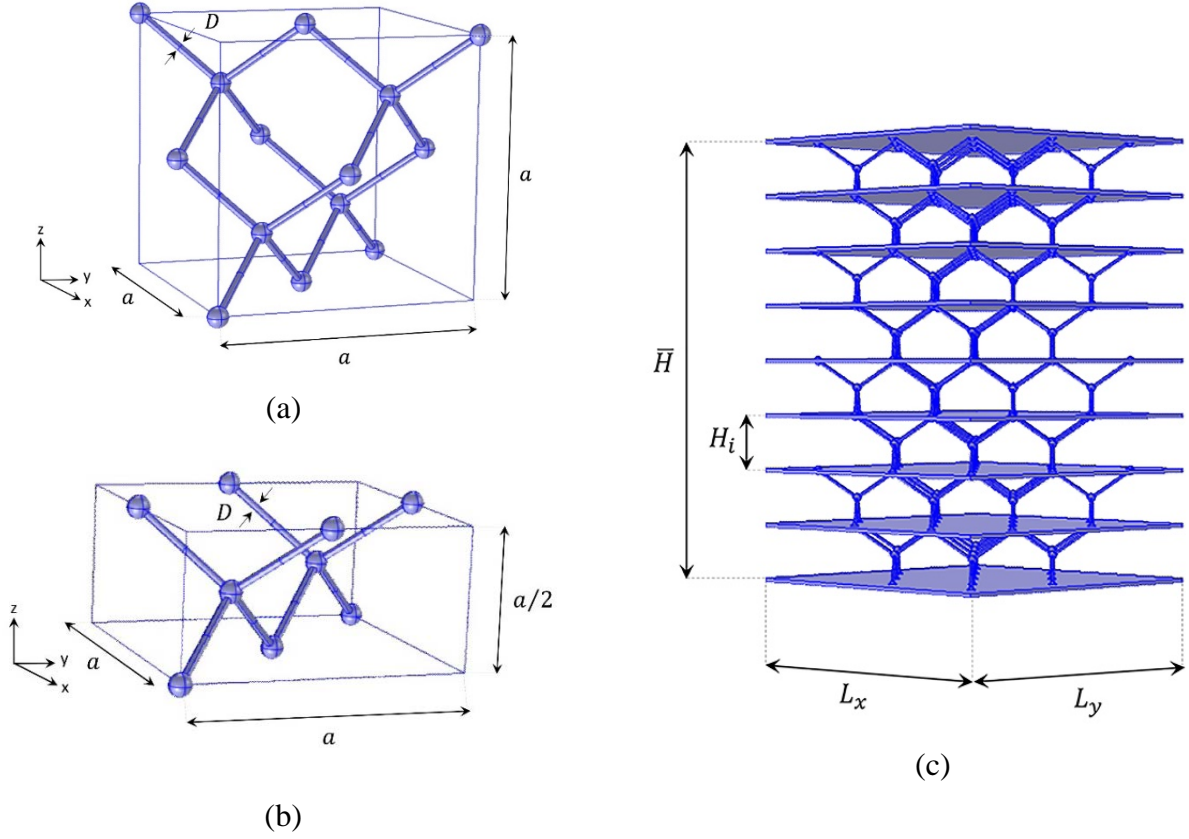

Figure 1. (a) Face-centered-cubic (fcc) unit cell of a pentamode lattice formed by four primitive unit cells. (b) Sub-lattice of the fcc unit cell formed by two primitive unit cells (sfcc cell). (c) Multilayered structure obtained by alternating sfcc lattices and stiffening plates.

## 3. Statics and kinematics of an elementary sfcc module

We begin by studying the static and kinematic problems of the elementary sfcc module formed by a primitive unit cell connected to end plates (Figure 2). We number the nodes forming such a module as shown in Figure 2, and sort the rods according to the following connection table: $\{(5-1),(5-2),(5-3),(5-4)\}$.

We take as reference the placement $\mathcal{B}$ such that the position vectors of the nodes are given by ("isotropic" pentamode placement, cf. [16])

$$\boldsymbol{n}_1 - \boldsymbol{n}_5 = \begin{bmatrix} a/4 \\ -a/4 \\ a/4 \end{bmatrix}, \quad \boldsymbol{n}_2 - \boldsymbol{n}_5 = \begin{bmatrix} -a/4 \\ a/4 \\ a/4 \end{bmatrix}, \quad \boldsymbol{n}_3 - \boldsymbol{n}_5 = \begin{bmatrix} -a/4 \\ -a/4 \\ -a/4 \end{bmatrix}, \quad \boldsymbol{n}_4 - \boldsymbol{n}_5 = \begin{bmatrix} a/4 \\ -a/4 \\ -a/4 \end{bmatrix} \qquad (1)$$

where $a$ denotes the lattice constant (Figure 1a); and $\boldsymbol{n}_5 = [x_5 \ y_5 \ z_5]^T$ denotes the position vector of the inner node, which we leave arbitrary. The next sections study an incremental motion of the elementary module from $\mathcal{B}$, by using a superimposed dot to denote incremental quantities related to such a motion, and linearizing the incremental equilibrium and compatibility equilibrium equations in the increments ("small" displacements from the reference configuration) [17].

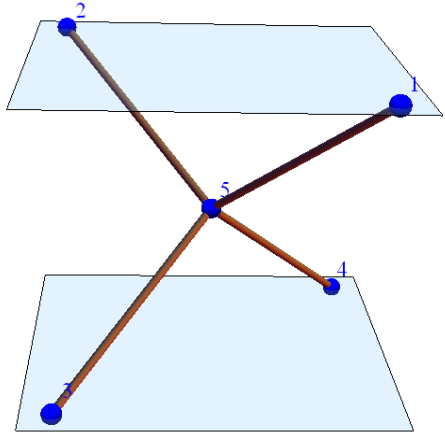

Figure 2. Elementary module of a sfcc system.

### 3.1 Incremental static problem

The incremental equilibrium equations of the elementary module in Figure 2 can be written in the following matrix form

$$A\dot{t} = \dot{f} \qquad (2)$$

Said $\mathbb{R}$ is the set of real numbers, here $\dot{t} \in \mathbb{R}^4$ is the vector of the incremental internal forces carried by the rods (incremental axial forces), $\dot{f} \in \mathbb{R}^{15}$ is the vector collecting the Cartesian components of the incremental external forces applied at the nodes (including the forces exerted by the stiffening plates), and $A$ is the equilibrium matrix. It is a simple task to verify that the latter has the following expression

$$A = \begin{bmatrix} \alpha & -\alpha & \alpha & 0 & 0 & 0 & 0 & 0 & 0 & 0 & 0 & 0 & -\alpha & \alpha & -\alpha \\ 0 & 0 & 0 & -\alpha & \alpha & \alpha & 0 & 0 & 0 & 0 & 0 & 0 & \alpha & -\alpha & -\alpha \\ 0 & 0 & 0 & 0 & 0 & 0 & -\alpha & -\alpha & -\alpha & 0 & 0 & 0 & \alpha & \alpha & \alpha \\ 0 & 0 & 0 & 0 & 0 & 0 & 0 & 0 & 0 & \alpha & \alpha & -\alpha & -\alpha & -\alpha & \alpha \end{bmatrix}^T \qquad (3)$$

where $T$ denotes the transposition symbol, and it results in $\alpha = \frac{1}{\sqrt{3}}$. In partitioned form, the algebraic system (2) can be written as follows

$$\begin{bmatrix} A_{11} & \vdots & A_{12} \\ \cdots & & \cdots \\ A_{21} & \vdots & A_{22} \end{bmatrix} \begin{bmatrix} \dot{t}_1 \\ \cdots \\ \dot{t}_2 \end{bmatrix} = \begin{bmatrix} \dot{f}_1 \\ \cdots \\ \dot{f}_2 \end{bmatrix} \qquad (4)$$

In Eqn. (4), $\dot{t}_1 \in \mathbb{R}^3$ is the vector of the incremental internal forces carried by rods $(5-1)$, $(5-2)$, and $(5-3)$; $\dot{t}_2 \in \mathbb{R}$ is a vector with a single entry equal to the incremental internal force carried by the rod $(5-4)$; $\dot{f}_1 = \mathbb{R}^3$ is the vector of the components of the incremental external forces acting on node 5; $\dot{f}_2 \in \mathbb{R}^{12}$ is the vector of the components' incremental external forces acting on nodes 1-4; and it results in

$$A_{11} = \begin{bmatrix} -.\alpha & \alpha & \alpha \\ \alpha & -\alpha & \alpha \\ -\alpha & -\alpha & \alpha \end{bmatrix}, \quad A_{12} = \begin{bmatrix} -\alpha \\ -\alpha \\ \alpha \end{bmatrix} \qquad (5)$$

$$\boldsymbol{A}_{21} = \begin{bmatrix} \alpha & -\alpha & \alpha & 0 & 0 & 0 & 0 & 0 & 0 & 0 & 0 & 0 \\ 0 & 0 & 0 & -\alpha & \alpha & \alpha & 0 & 0 & 0 & 0 & 0 & 0 \\ 0 & 0 & 0 & 0 & 0 & 0 & -\alpha & -\alpha & -\alpha & 0 & 0 & 0 \end{bmatrix}^T, \tag{6}$$

$$\boldsymbol{A}_{22} = \begin{bmatrix} 0 & 0 & 0 & 0 & 0 & 0 & 0 & 0 & 0 & \alpha & \alpha & -\alpha \end{bmatrix}^T \tag{7}$$

It is easy to verify that $\boldsymbol{A}_{11}$ is invertible. Assuming $\dot{\boldsymbol{f}}_1 = 0$, i.e., supposing that the internal node 5 is unloaded, we easily get

$$\dot{\boldsymbol{t}}_1 = -\boldsymbol{A}_{11}^{-1}\boldsymbol{A}_{12}\dot{\boldsymbol{t}}_2 = \begin{bmatrix} \dot{t}_2 \\ \dot{t}_2 \\ \dot{t}_2 \end{bmatrix} \tag{8}$$

(8) shows that all the rods forming the elementary sfcc module carry equal axial forces, under the assumption that the inner nodes are unloaded.

### 3.2 Incremental kinematic problem

We now let $\dot{\boldsymbol{d}}_1 = \mathbb{R}^3$ denote the vector collecting the Cartesian components of the incremental displacement of node 5, and let $\dot{\boldsymbol{d}}_2 \in \mathbb{R}^{12}$ indicate the vector of the components of the incremental displacements of nodes 1-4. In addition, we let $\dot{\boldsymbol{e}}_1 \in \mathbb{R}^3$ denote the vector of the incremental elongations of the rods $(5-1)$, $(5-2)$, and $(5-3)$, and let $\dot{\boldsymbol{e}}_2 \in \mathbb{R}$ denote a vector with a single entry equal to the incremental elongation of the rod $(5-4)$. We assume that the incremental elongation $\dot{e}$ of the generic rod is related to the corresponding incremental axial force $\dot{t}$ through the following elastic constitutive equation

$$\dot{t} = \frac{E_0 \dot{e}}{\ell} \tag{9}$$

$\ell = \sqrt{3}a/4$ denoting the length of the rod in the reference placement $\mathcal{B}$.

The incremental displacements and elongations of the elementary module are related to each other through the following incremental kinematic problem

$$\begin{bmatrix} \boldsymbol{B}_{11} & \vdots & \boldsymbol{B}_{12} \\ \cdots & & \cdots \\ \boldsymbol{B}_{21} & \vdots & \boldsymbol{B}_{22} \end{bmatrix} \begin{bmatrix} \dot{\boldsymbol{d}}_1 \\ \cdots \\ \dot{\boldsymbol{d}}_2 \end{bmatrix} = \begin{bmatrix} \dot{\boldsymbol{e}}_1 \\ \cdots \\ \dot{\boldsymbol{e}}_2 \end{bmatrix} \tag{10}$$

where

$$\boldsymbol{B}_{11} = \boldsymbol{A}_{11}^T = \begin{bmatrix} -\alpha & \alpha & -\alpha \\ \alpha & -\alpha & -\alpha \\ \alpha & \alpha & \alpha \end{bmatrix}, \quad \boldsymbol{B}_{12} = \boldsymbol{A}_{21}^T = \begin{bmatrix} \alpha & -\alpha & \alpha & 0 & 0 & 0 & 0 & 0 & 0 & 0 & 0 & 0 \\ 0 & 0 & 0 & -\alpha & \alpha & \alpha & 0 & 0 & 0 & 0 & 0 & 0 \\ 0 & 0 & 0 & 0 & 0 & 0 & -\alpha & -\alpha & -\alpha & 0 & 0 & 0 \end{bmatrix} \tag{11}$$

$$\boldsymbol{B}_{21} = \boldsymbol{A}_{12}^T = [-\alpha \quad -\alpha \quad \alpha], \quad \boldsymbol{B}_{22} = \boldsymbol{A}_{22}^T = [0 \quad 0 \quad 0 \quad 0 \quad 0 \quad 0 \quad 0 \quad 0 \quad 0 \quad \alpha \quad \alpha \quad -\alpha] \quad (12)$$

Due to the assumption of rigid behavior of the terminal pates, and without loss of generality, we hereafter assume that the bottom plate is at rest, and the motion of the top plate can be represented as the composition of an incremental translation $\dot{\boldsymbol{v}}$ and an infinitesimal incremental rotation with axial vector $\dot{\boldsymbol{\varphi}}$, about its center of mass $G_t$. Under such assumptions, $\dot{\boldsymbol{d}}_2 = \dot{\boldsymbol{d}}_2(\dot{\boldsymbol{v}}, \dot{\boldsymbol{\varphi}})$ in Eqn. (10) describes a relative rigid motion of the terminal bases of the elementary module. By solving Eqn. (10) for $\dot{\boldsymbol{e}}_2$, we get

$$\dot{\boldsymbol{e}}_2 = \boldsymbol{B}_{21}\dot{\boldsymbol{d}}_1 + \boldsymbol{B}_{22}\dot{\boldsymbol{d}}_2 \quad (13)$$

Taking into account Eqn. (13), the constitutive assumption (9) and Eqn. (8), we conclude that all the rods of the elementary module exhibit equal incremental elongations, and it results

$$\dot{\boldsymbol{e}}_1 = \dot{\boldsymbol{e}}_2 \mathbf{1} = (\boldsymbol{B}_{21}\dot{\boldsymbol{d}}_1 + \boldsymbol{B}_{22}\dot{\boldsymbol{d}}_2)\mathbf{1} \quad (14)$$

$\mathbf{1}$ denoting the vector of $\mathbb{R}^3$ with all the entries equal to one. Solving now Eqn. (10) for $\dot{\boldsymbol{d}}_1$, and taking into account Eqn. (14), we get

$$\dot{\boldsymbol{d}}_1 = \widehat{\boldsymbol{B}}_{11}^{-1}[(\boldsymbol{B}_{22}\dot{\boldsymbol{d}}_2)\mathbf{1} - \boldsymbol{B}_{12}\dot{\boldsymbol{d}}_2] \quad (15)$$

where

$$\widehat{\boldsymbol{B}}_{11} = \begin{bmatrix} 0 & 2\alpha & -2\alpha \\ 2\alpha & 0 & -2\alpha \\ 2\alpha & 2\alpha & 0 \end{bmatrix}, \quad \widehat{\boldsymbol{B}}_{11}^{-1} = \begin{bmatrix} -\beta & \beta & \beta \\ \beta & -\beta & \beta \\ -\beta & -\beta & \beta \end{bmatrix}, \quad (16)$$

$$\beta = \frac{1}{4\alpha} = \frac{\sqrt{3}}{4} \quad (17)$$

It is worth noting that it results

$$\widehat{\boldsymbol{B}}_{11} = \boldsymbol{B}_{11} - \widehat{\boldsymbol{B}}_{21} \quad (18)$$

$\widehat{\boldsymbol{B}}_{21}$ denoting the $3 \times 3$ matrix having each row equal to $\boldsymbol{B}_{21}$. Eqn. (15) allows us to compute the incremental displacement $\dot{\boldsymbol{d}}_1 = \dot{\boldsymbol{d}}_1(\dot{\boldsymbol{v}}, \dot{\boldsymbol{\varphi}})$ of the inner node of the elementary module, which corresponds to any arbitrary relative rigid motion $\dot{\boldsymbol{d}}_2(\dot{\boldsymbol{v}}, \dot{\boldsymbol{\varphi}})$ of the terminal bases.

Let $\dot{\boldsymbol{d}}(\dot{\boldsymbol{v}}, \dot{\boldsymbol{\varphi}}) = [\dot{\boldsymbol{d}}_1^T(\dot{\boldsymbol{v}}, \dot{\boldsymbol{\varphi}}) \vdots \dot{\boldsymbol{d}}_2^T(\dot{\boldsymbol{v}}, \dot{\boldsymbol{\varphi}})]^T$ denote the overall displacement vector of the elementary module associated with a given $\dot{\boldsymbol{d}}_2(\dot{\boldsymbol{v}}, \dot{\boldsymbol{\varphi}})$ through Eqn. (15). When a relative rigid motion of the terminal bases is such that Eqns. (13)-(14) return $\dot{e} = 0$ in each rod, i.e., $\dot{\boldsymbol{e}}_2 = 0$ and $\dot{\boldsymbol{e}}_1 = \boldsymbol{0}$, we say that such a motion represents an *infinitesimal mechanism* of the elementary module from the reference placement $\mathcal{B}$.

## 4. Infinitesimal mechanisms

The present section studies the relative motions of the end plates of a monolayer sfcc system that produce infinitesimal mechanisms (zero-energy modes) of the system from the reference placement $\mathcal{B}$. We will see that such mechanisms are generated by relative horizontal displacements (*shear mechanisms*), and the *twisting* of the end plates. This differs from the behavior of a confined fcc system, i.e., a mono- or multi-layer structure that makes use of the fcc unit cell shown in Figure 1a. It is not difficult to prove that a structure formed by the fcc layer and stiffening plates, when equipped with hinged connections, exhibits finite mechanisms for each possible relative motion of the end plates. Such a result implies that confined fcc systems are unable to carry vertical and bending loads in the pure stretching regime.

### 4.1 Shear mechanisms

An infinitesimal shear mechanism along the $x$ axis of a monolayer sfcc system is obtained by imposing an incremental $x$-translation of amplitude $\dot{u}$ to the top plate and keeping the bottom plate at rest. With reference to the elementary module in Figure 2, such a relative rigid motion of the end plates is described by

$$\dot{\boldsymbol{d}}_2 = [[\dot{u},0,0]^T \quad [\dot{u},0,0]^T \quad [0,0,0]^T \quad [0,0,0]^T]^T \tag{19}$$

and it is easily shown that it produces zero incremental elongations in the rods, and the following incremental displacement of the inner node

$$\dot{\boldsymbol{d}}_1 = [\dot{u}/2 \quad -\dot{u}/2 \quad 0]^T \tag{20}$$

via the Eqns. (13)-(15) of Sect.3.2.

Similarly, an infinitesimal shear mechanism of the elementary module along the $y$ axis, which is induced by the following relative rigid motion of the end plates

$$\dot{\boldsymbol{d}}_2 = [[0,\dot{v},0]^T \quad [0,\dot{v},0]^T \quad [0,0,0]^T \quad [0,0,0]^T]^T \tag{21}$$

produces zero incremental elongations in the rods, and

$$\dot{\boldsymbol{d}}_1 = [-\dot{v}/2 \quad \dot{v}/2 \quad 0]^T \tag{22}$$

We show in Figure 3 the $x$- and $y$-shear mechanisms of a sfcc structure formed by the assembly of eight elementary modules.

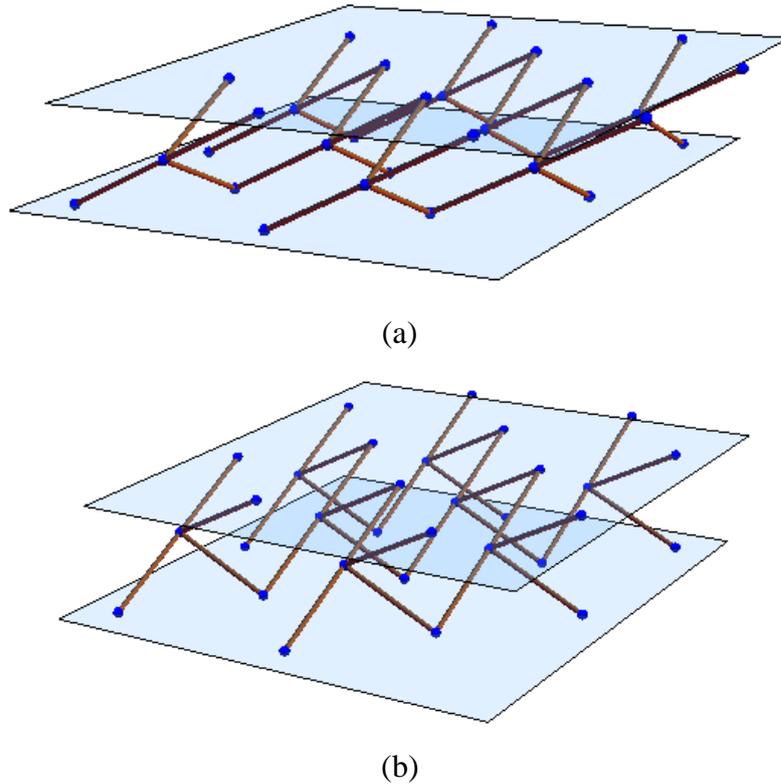

(a)

(b)

Figure 3. Shear mechanisms of a monolayer sfcc system: (a) $x$-axis shear mechanism; (b) $y$- axis shear mechanism (the infinitesimal displacements have been amplified for visual clarity).

## 4.2 Twisting mechanism

We now consider an incremental infinitesimal rotation $\dot{\varphi}_z$ of the top plate about the $z$- axis, again keeping the bottom plate at rest. With reference to the elementary module in Figure 2, such a relatively rigid motion of the end plates is described by

$$\dot{\boldsymbol{d}}_2 = \begin{bmatrix} -\dot{\varphi}_z\left(-\dfrac{a}{4} + y_5 - y_G\right) \\ \dot{\varphi}_z\left(\dfrac{a}{4} + x_5 - x_G\right) \\ 0 \\ -\dot{\varphi}_z\left(\dfrac{a}{4} + y_5 - y_G\right) \\ \dot{\varphi}_z\left(\dfrac{a}{4} + x_5 - x_G\right) \\ 0 \\ 0 \\ 0 \\ 0 \\ 0 \\ 0 \\ 0 \end{bmatrix} \tag{23}$$

where $x_G$ and $y_G$ denote the $x$ and $y$ coordinates of $G_t$, respectively. On using Eqns. (13)-(15) of Sect.3.2, it is easy to verify that the twisting of the end plates produces zero incremental elongations in the rods, and the following incremental displacement of the inner node

$$\dot{\boldsymbol{d}}_1 = \begin{bmatrix} -\dfrac{\dot{\varphi}_z}{2}(y_5 - y_G) + \dfrac{\dot{\varphi}_z}{2}(x_5 - x_G) \\ \dfrac{\dot{\varphi}_z}{2}(y_5 - y_G) - \dfrac{\dot{\varphi}_z}{2}(x_5 - x_G) \\ 0 \end{bmatrix} \tag{24}$$

A graphical illustration of the twisting mechanism is provided in Figure 4.

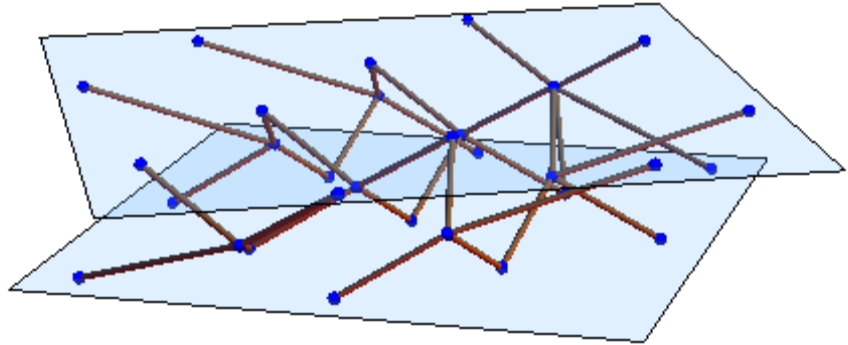

Figure 4. Twisting mechanisms of a monolayer sfcc system (the infinitesimal displacements have been amplified for visual clarity).

## 5. Infinitesimal elastic deformations

Here we examine the relative motions of the terminal plates of an sfcc layer that produce incremental elastic deformations of the system from the reference placement $\mathcal{B}$.

### 5.1 Vertical deformation

Let us impress an incremental vertical translation of amplitude $\dot{w}$ to the top plate by keeping the bottom plate at rest. In correspondence with the elementary module in Figure 2, such an incremental deformation corresponds to assuming

$$\dot{\boldsymbol{d}}_2 = [[0,0,\dot{w}]^T \quad [0,0,\dot{w}]^T \quad [0,0,0]^T \quad [0,0,0]^T]^T \tag{25}$$

From Eqn. (15) of Sect. 3.2 we deduce in the present case

$$\dot{\boldsymbol{d}}_1 = [0 \quad 0 \quad \dot{w}/2]^T \tag{26}$$

On the other hand, Eqns. (13)-(14) lead us to recognize that the current deformation mode induces the following incremental elongations and incremental axial forces in all the rods of the system

$$\dot{e} = \frac{\alpha \dot{w}}{2} = \frac{\dot{w}}{2\sqrt{3}} \tag{27}$$

$$\dot{t} = \frac{2E_0 s}{3a} \dot{w} \tag{28}$$

It is worth noting that the vertical component of $\dot{t}$ is given by

$$\dot{f}_v = \alpha \dot{t} = \frac{2E_0 s}{3\sqrt{3}a} \dot{w} \tag{29}$$

Taking into account that the sfcc unit cell includes four rods attached to the top plate (cf. Figure 1b), we now compute the total incremental vertical force carried by an sfcc system with $n_x \times n_y$ unit cells in the horizontal plane (Figure 5), which is given by

$$\dot{F}_v = 4 n_x n_y \dot{f}_v = \frac{8 E_0 s n_x n_y}{3\sqrt{3}a} \dot{w} \tag{30}$$

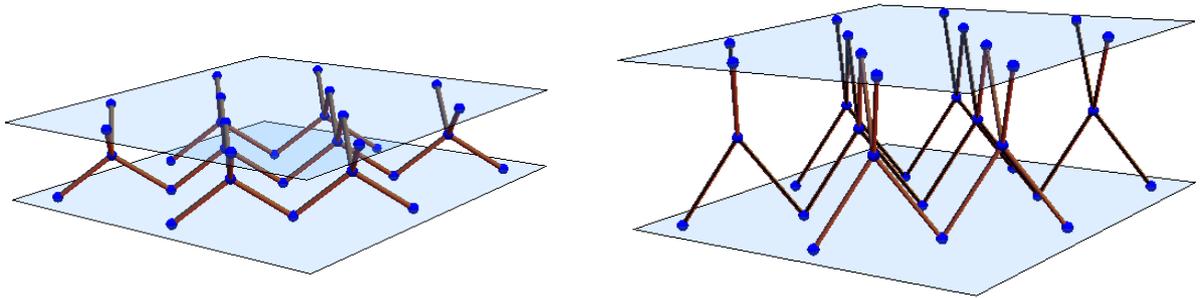

Figure 5. Vertical deformation of a monolayer sfcc system (left: undeformed configuration; right: deformed configuration – the infinitesimal displacements have been amplified for visual clarity).

Making use of Eqn.(30), we compute the incremental vertical stiffness of the system as follows

$$K_v = \frac{\dot{F}_v}{\dot{w}} = \frac{8E_0 s n_x n_y}{3\sqrt{3}a} \tag{31}$$

It is easily observed that such a quantity grows linearly with the rods' Young modulus $E_0$, the numbers of unit cells placed along the $x$ and $y$ axes, and the rods' cross-section area $s$. $K_v$ is instead inversely proportional to the lattice constant $a$.

Introducing now the solid volume fraction of the unit cell, defined as follows (cf. Figure 1a)

$$\phi = \frac{8s\ell}{a \times a \times \frac{a}{2}} = \frac{4\sqrt{3}s}{a^2} \tag{32}$$

we can rewrite Eqn. (31) in the form

$$\frac{K_v}{E_0 a} = \frac{2}{9} n_x n_y \phi \tag{33}$$

Eqn. (33) highlights that $K_v$ varies linearly with $\phi$, as is graphically shown in the plot of Figure 6, for the case of a square system featuring $n_x = n_y = n_a$. Such a plot explicitly reports the numerical values of the dimensionless quantity $K_v/(E_0 a)$ that correspond to $\phi = 3\%$ and varying values of $n_a$.

The effective compression modulus $E_c$ of the sfcc system is defined as follows

$$E_c = \frac{K_v \frac{a}{2}}{n_x n_y a^2} = \frac{4 E_0 s}{3\sqrt{3} a^2} = \frac{\phi}{9} E_0 \tag{34}$$

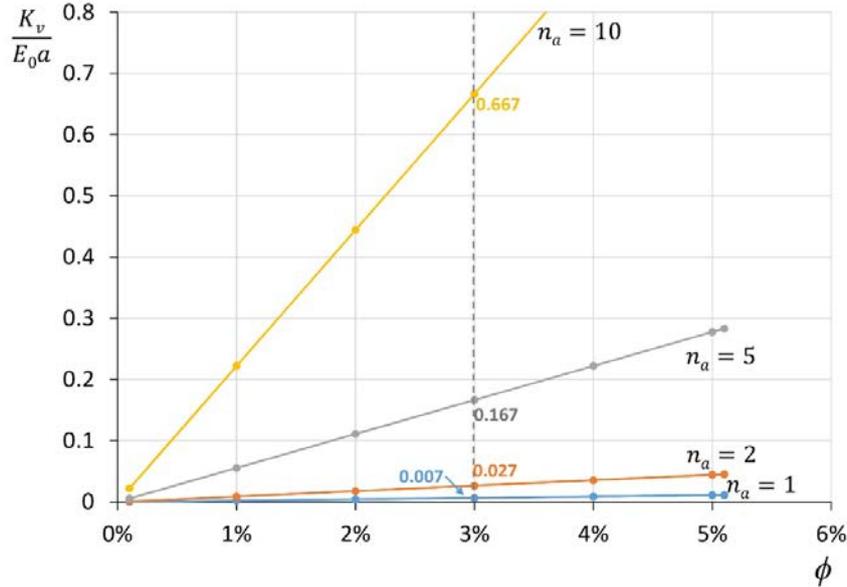

Figure 6. Vertical stiffness of a monolayer sfcc system vs. the solid volume fraction, for different numbers of unit cells in the horizontal plane ($\boldsymbol{n_a = n_x = n_y}$).

It is interesting to note that the Young modulus of an unconfined pentamode lattice is zero in the stretch-dominated limit [4], while Eqn. (34) predicts an effective compression modulus equal to 2/3 of the Young modulus of the stiffest isotropic elastic networks analyzed in [14], for a confined sfcc system. Eqn. (34) also shows that $E_c$ increases with the rods' cross-section area $s$, and decreases with the lattice constant $a$, being a linear function of the solid volume fraction $\phi$ (stretch-dominated response, cf., e.g., [14][18][19]).

The compression modulus $E_c$ of an elastomeric layer confined between stiffening plates, which is commonly employed to form rubber seismic isolation systems known as rubber bearings, is controlled by a shape factor $S$ which is defined as the ratio between the load area and the force-free (lateral) area [20]-[23]. Such a quantity is hence directly proportional to a characteristic dimension of the load area, and inversely proportional to the rubber pad thickness (see, e.g., [20]). We observe from Eqn. (34) that the compression modulus of a "pentamode bearing", formed by a sfcc lattice confined between stiffening plates, is inversely proportional to the lattice thickness, which is indeed equal to $a/2$ (Figure 1b). It is also worth noting that the role played by the characteristic transverse dimension of the rubber pads in rubber bearings is replaced by the cross-section area of the lattice rods in a pentamode bearing.

## 5.2 Bending deformation

A bending deformation about the $y$- axis of a monolayer sfcc system is obtained by imposing an incremental infinitesimal rotation $\dot\varphi_y$ to the top plate and keeping the bottom plate at rest. When applied to the elementary module in Figure 2, such an incremental deformation corresponds to assuming

$$\dot{\boldsymbol{d}}_2 = \left[\left[0,0,-\dot\varphi_y\left(\frac{a}{4}+\Delta x_G\right)\right]^T \quad \left[0,0,-\dot\varphi_y\left(-\frac{a}{4}+\Delta x_G\right)\right]^T \quad [0,0,0]^T \quad [0,0,0]^T\right]^T \tag{35}$$

From Eqns. (15) and (13)-(14), we deduce that, in the present case

$$\dot{\boldsymbol{d}}_1 = \begin{bmatrix} -\dfrac{a}{8}\dot\varphi_y \\ \dfrac{a}{8}\dot\varphi_y \\ -\dfrac{\dot\varphi_y}{2}\Delta x_G \end{bmatrix} \tag{36}$$

$$\dot e = \alpha\left(-\frac{a}{8}\dot\varphi_y\right) + \alpha\left(\frac{a}{8}\dot\varphi_y\right) - \alpha\left(-\frac{\dot\varphi_y}{2}\Delta x_G\right) = \frac{\alpha \Delta x_G}{2}\dot\varphi_y \tag{37}$$

$$\dot t = \frac{E_o s \dot e}{\ell} = \frac{2 E_o s}{3 a}\dot\varphi_y \Delta x_G \tag{38}$$

where $\dot e$ denotes the incremental elongation in the generic rod of the elementary module; $\dot t$ denotes the incremental axial force carried by the generic rod; and it results in: $\Delta x_G = x_5 - x_G$. The vertical component of $\dot t$ is given by

$$\hat{\dot f}_v = \alpha \dot t = \frac{2 E_0 s}{3\sqrt{3}a}\dot\varphi_y \Delta x_G \tag{39}$$

We now examine a monolayer sfcc system composed of $n_x$ unit cells along the $x-$axis and $n_y$ unit cells along the $y-$ axis (Figure 7). On considering that each elementary module of such a system shows two rods attached to the top plate (Figure 1a), we compute the total incremental bending moment $\dot M_y$ carried by the system as follows

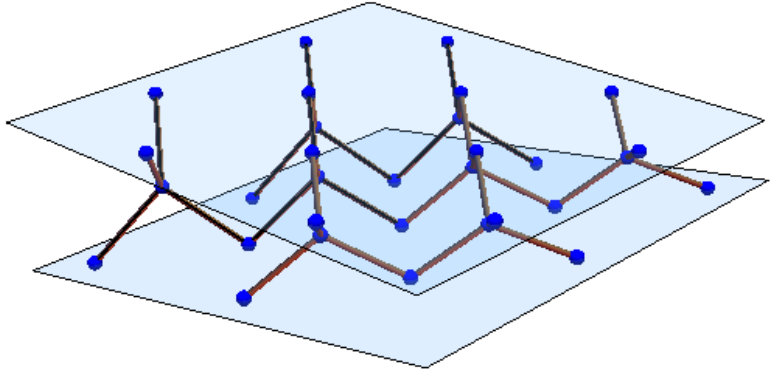

Figure 7. Bending deformation about the $y$-axis of a monolayer sfcc system (the infinitesimal displacements have been amplified for visual clarity).

$$\dot{M}_y = \frac{2E_0 s}{3\sqrt{3}a} \dot{\varphi}_y \sum_{i=1}^{n_{mod}} \left[\left(\Delta x_{G_i} - \frac{a}{4}\right) + \left(\Delta x_{G_i} + \frac{a}{4}\right)\right] \Delta x_{G_i} \tag{40}$$

where $n_{mod}$ indicates the total number of elementary modules forming the system (two modules for each unit cell, cf. Figure 1b), and $\Delta x_{G_i}$ denotes the relative $x-$ coordinate of the central node of the $i$-th module with respect to $G_t$. Upon numbering the elementary modules from left (negative $x$) to right (positive $x$), we get

$$\Delta x_{G_i} = -n_x \frac{a}{2} + \frac{a}{4} + (i-1)\frac{a}{2} = \frac{a}{4}[2(n_x + i - 1) - 1] \tag{41}$$

Making use of Eqns. (40)-(41), we finally obtain

$$K_{\varphi_y} = \frac{2E_0 s}{3\sqrt{3}} \sum_{i=1}^{n_{mod}} 2\Delta x_{G_i}^2 = \frac{4E_0 s}{3\sqrt{3}a} n_y \sum_{i=1}^{2n_x} \left\{\frac{a}{4}[2(n_x + i - 1) - 1]\right\}^2 = \frac{E_0 s\, a\, n_x(4n_x^2 - 1)n_y}{18\sqrt{3}} \tag{42}$$

From Eqns. (32) and (42), we get

$$\frac{K_{\varphi_y}}{E_0 a^3} = \frac{1}{216} n_x(4n_x^2 - 1)n_y \phi \tag{43}$$

Figure 8 plots the dimensionless quantity $K_{\varphi_y}/(E_0 a^3)$ against the lattice solid volume fraction $\phi$, in the case of a square system ($n_x = n_y = n_a$). Eqns. (42)-(43) show that $K_{\varphi_y}$ grows with the lattice constant $a$, the rods' axial stiffness $E_0 s$, the number of unit cells placed along the $x$ and $y$ axes, and the lattice solid volume fraction $\phi$. In particular, $K_{\varphi_y}$ grows linearly with the number of unit cells in the $y$-direction, and quadratically with the number of unit cells in the $x$- direction. It is also worth remarking that such a quantity grows with the layer thickness ($a/2$), as opposed to the vertical stiffness $K_v$, which is instead inversely proportional to $a$ (cf. the previous section). Analytic formulae for the bending stiffness $K_{\varphi_x}$ about the $x-$axis (Figure 9) are easily obtained by switching $n_x$ with $n_y$ in Eqns. (42)-(43).

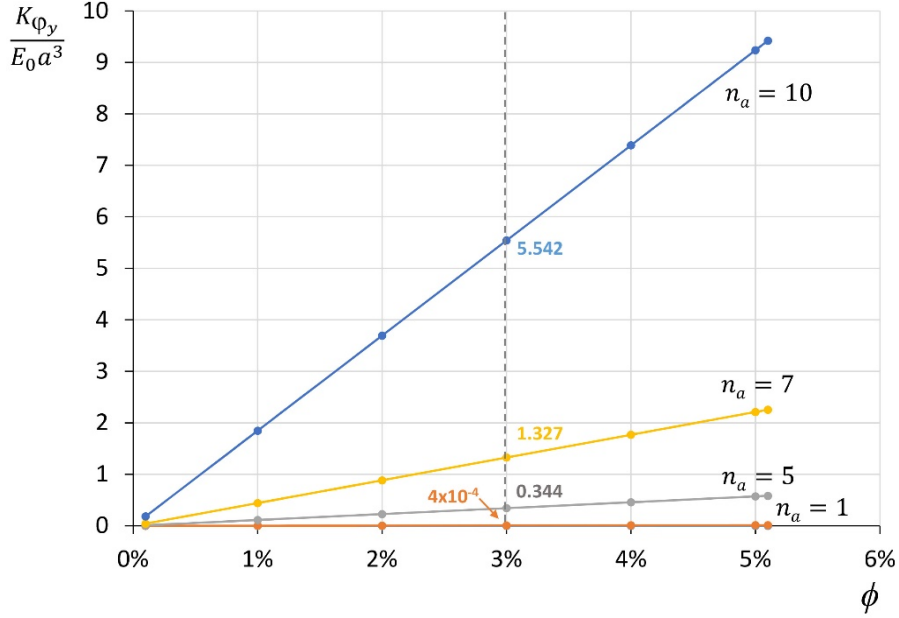

Figure 8. Bending stiffness about the $y-$axis of a monolayer sfcc system vs. the solid volume fraction, for different numbers of unit cells in the horizontal plane ($n_a = n_x = n_y$).

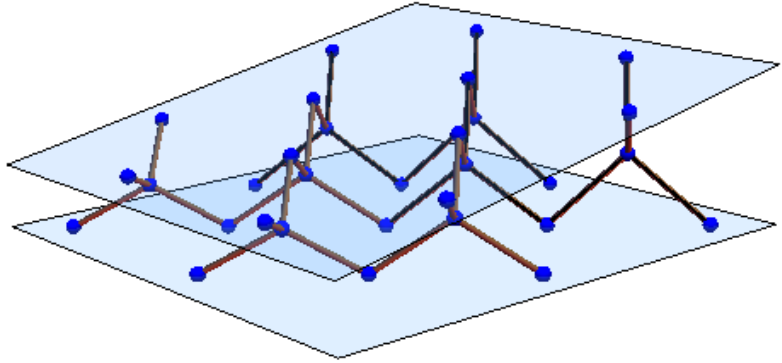

Figure 9. Bending deformation about the $x$- axis of a monolayer sfcc system (the infinitesimal displacements have been amplified for visual clarity).

### 5.3 Multi-layer systems

Let us now examine multi-layer sfcc systems formed by alternating a number $n_z$ of sfcc layers and confinement plates. The vertical and bending stiffness properties of such systems are easily obtained, on assuming that the layers forming the laminated structure are connected in series. We get

$$K_v = \frac{1}{\sum_{i=1}^{n_z}\frac{1}{K_{v_i}}}; \quad K_\varphi = \frac{1}{\sum_{i=1}^{n_z}\frac{1}{K_{\varphi_i}}} \tag{44}$$

where $K_{v_i}$ and $K_{\varphi_i}$ denote the vertical stiffness and the bending stiffness (about either the $x-$ or the $y-$ axis) of the $i$-th layer. On assuming that $K_{v_i}$ and $K_{\varphi_i}$ are constant from layer to layer, we obtain

$$E_c = \frac{K_v H}{A} = \frac{H}{A}\frac{1}{\frac{n_z}{K_{v_i}}} = \frac{K_{v_i} H_i}{A} = E_{c_i} = \frac{4E_o s}{3\sqrt{3}a^2} \tag{45}$$

where
$$A = n_x n_y a^2 \qquad (46)$$

denotes the area of the stiffening plates covered by the pentamode lattices ("load area"), and $E_{c_i}$ denotes the effective compression modulus of the generic layer. Eqn. (45) shows that the compression modulus of the layered system is equal to that of each individual layer, under the above assumptions. We now focus attention on a square multi-layer system ($n_x = n_y = n_a$), observing that in such a case it results in

$$a = \frac{L}{n_a} = \frac{2H}{n_z} \qquad (47)$$

where $L = L_x = L_y$ denotes the edge-length of the load area. The use of Eqn. (47) into Eqn. (45) leads us to the following expressions of the vertical stiffness

$$K_v = \frac{2}{3\sqrt{3}} \frac{E_0 s L}{H^2} n_a n_z \qquad (48)$$

and the effective compression modulus

$$E_c = \frac{2}{3\sqrt{3}} \frac{E_0 s}{LH} n_a n_z \qquad (49)$$

of the multi-layer system under consideration. For fixed values of $L, H, E_0$ and $s$, Eqn. (49) shows that $K_v$ and $E_c$ of such a system scale linearly with both $n_a$ and $n_z$ (cf. Figure 10). By keeping the above variables fixed, it is easy to realize, e.g., that $K_v$ and $E_c$ get four times larger when doubling the number of cells in the horizontal plane and the number of layers. It is worth noting that, when doubling $n_z$ and keeping $H$ fixed, Eqn. (47) implies that one needs to halve the lattice constant $a$. On the other hand, the same equation implies that, in the same conditions, one contemporarily needs to double $n_a$, in order to keep also $L$ as constant.

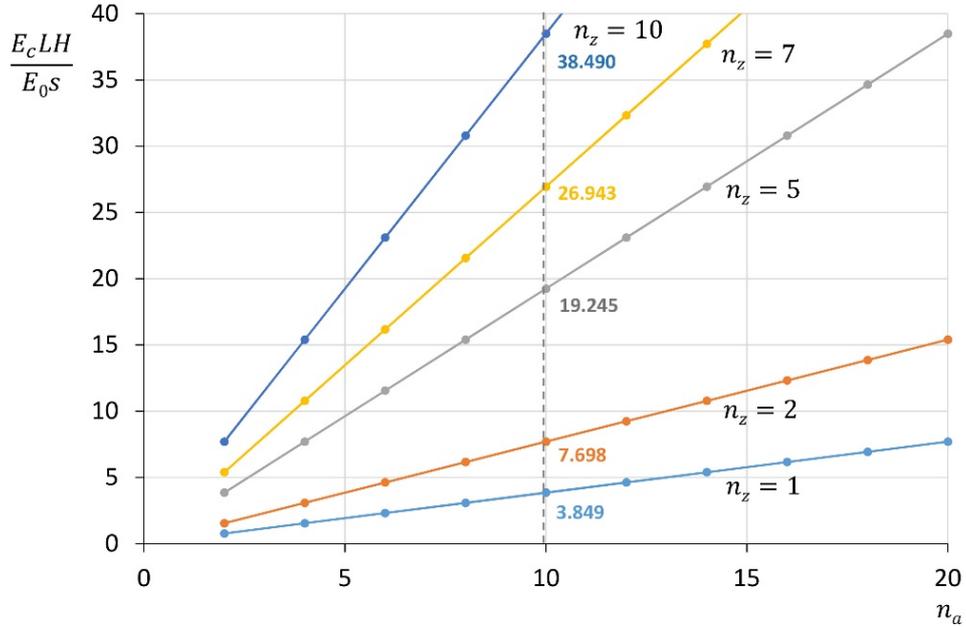

Figure 10. Effective compression modulus of a multilayer sfcc system vs. the number of unit cells in the horizontal plane ($\boldsymbol{n_a = n_x = n_y}$) and the number of layers ($\boldsymbol{n_z}$).

## 6. Numerical results

The present section numerically investigates the elastic response of physical models of sfcc pentamode bearings, making use of steel bars grade S335JH, with $E_o = 210$ GPa [24]. The examined systems can be actually built using rods and ball joints commonly employed for the realization of space grids [15]. Our systems feature $2 \times 2$ sfcc unit cells on the horizontal plane ($n_x = n_y = 2$), lattice constant $a = 1200$ mm, layer height 600 mm, stiffening plate edge length $L = 2400$ mm, and hollow circular rods with length $\ell = 519.6$ mm, 48.3 mm diameter and 5 mm wall thickness [24]. Their assembly would require ball joints with diameters of about 70-90 mm, i.e., joints with minimal dimensions among those commonly employed for space grids [15].

The elastic response of the systems under consideration is analyzed through the commercial software Sap2000®, making use of a finite element model (FEM) equipped with beam (frame) elements to describe the rods, perfectly hinged connections, and 2D rigid elements to describe the stiffening plates [25]. Table 1 compares finite element values of the vertical stiffness $K_v$ and the bending stiffness $K_\varphi = K_{\varphi_x} = K_{\varphi_y}$ of the examined systems with the theoretical previsions of the same quantities obtained through Eqns. (31), (42) and (44), for varying numbers of layers $n_z$. We observe an excellent matching between theoretical and FEM results, with maximum theory-FEM mismatch equal to $\sim 1.0\%$ ($K_\varphi$ of the single-layer system). It is worth noting that the $K_v$ and $K_\varphi$ coefficients of the models equipped with $n_z$ layers are approximatively equal to $1/n_z$ of those competing to the monolayer system, in line with the theory presented in Sect. 5.3.

Table 1. FEM predictions of the vertical stiffness $K_v$ [kN/mm] and the bending stiffness $K_\varphi$ [kNmm] of physical models of layered sfcc systems vs. theoretical values (TH).

| $n_z$ | $K_{v,FEM}$ | $K_{v,TH}$ | $K_{\varphi,FEM}$ | $K_{\varphi,TH}$ |
|---|---|---|---|---|
| 1 | 7.33E+02 | 7.33E+02 | 3.34E+08 | 3.300+08 |
| 2 | 3.66E+02 | 3.66E+02 | 1.65E+08 | 1.65E+08 |
| 3 | 2.43E+02 | 2.44E+02 | 1.10E+08 | 1.10E+08 |

## 7. Concluding remarks

We have presented a new class of mechanical metamaterials obtained by stiffening sfcc pentamode layers with confinement plates. Different from pentamode metamaterials that fill the Euclidean space with fcc unit cells formed by four primitive cells [1]-[12], the lattices analyzed in the present study make use of a sub-lattice of the pentamode fcc cell formed by only two primitive cells. Such a sfcc unit cell is repeated an arbitrary number of times in the horizontal plane, and is alternated to stiffening plates along the vertical axis. We have demonstrated that sfcc lattices feature only three zero-energy modes in the small strain, stretch-dominated regime, and exhibit finite (non-zero) stiffness against vertical loads and bending moments (Sects. 4-5). In such a regime, we have shown that they achieve an effective compression modulus equal to $2/3$ of the Young modulus of the stiffest elastic networks analyzed in Ref. [14]. This is a noteworthy result, since it is known that many cell fcc lattices with hinged connections instead exhibit zero Young modulus [4]. The finite element results provided in Sect. 6 allowed us to validate the analytic results presented in Sect. 5 with reference to the stiffness coefficients of single-layer and multi-layer sfcc structures.

Overall, we may conclude that the analyzed pentamode lattices can be effectively employed as novel impact protection gears and seismic isolation devices, by suitably designing the lattice geometry, the stiffness properties of the joints, and the lamination scheme, as a function of the operating conditions. Systems endowed with hinged or semi-rigid connections may be effective as impact protectors under impulsive shear loading. Recent research has revealed that that strongly nonlinear wave propagation in periodic media can be a feasible and convenient alternative to present state-of-the-art impact protection engineering [26]-[32]. Shear waves are particularly dangerous in many impact situations, and may lead to diffuse axonal injury in traumatic brain injuries induced by angular accelerations and decelerations of the head (refer, e.g, to the review paper [33] and references therein). The impact-absorbing liner of next-generation helmets could be designed to reproduce the skull-brain system, with the outer section mimicking the skull, through a pressure-wave mitigation lattice [32], and the inner section mimicking the cerebrospinal fluid, via micro- or small-scale confined pentamode metamaterials acting as "metafluid" lattices [11]. Macroscale sfcc systems with pinned or semi-rigid joints can also serve as next-generation seismic isolators, whose isolation properties may be finely adjusted to the structure being isolated [12]-[13]. Previous studies have pointed out several mechanical analogies between the bending-dominated response of confined pentamode lattices and the mechanics of seismic isolation devices alternating rubber layers and stiffening plates [20]-[23]. In both cases, the plates forming such laminated structures stiffen the compressive deformation mode of the system, and, at the same time, keep its compliance against shear actions sufficiently large [10]-[13], [20]-[23]. The outcomes of the present research allow us to extend the above findings to the case of the pure stretching response of sfcc pentamode metamaterials.

In closing, we point out a number of aspects of the present work that suggest directions for future research. Challenging extensions and generalizations of the current research regard the analytical modeling of the effective stiffness properties of confined pentamode lattices in the large strain regime, with special focus on the link between vertical and lateral stiffness properties. Another interesting task regards the design of systems that use hard materials for the stiffening plates, and soft materials for the bars ( e.g., nylon, PMMA, etc.). An optimal design of such systems may lead to extremely low shear moduli and a sufficiently high compression modulus. Additional extensions concern the computational modeling of the bending response and the dynamical behavior of confined pentamode lattices, allowing for damping, fracture, damage, and plasticity effects under large displacements. These modeling tasks need to be accompanied by an experimental characterization phase, with the aim of implementing and verifying the theoretical predictions. We plan to fabricate physical models of sfcc pentamode lattices at different scales, as well as employing additive manufacturing technologies at the micro-/small-scale [10], and space grid systems equipped with ball-joints at the macro-/large-scale [15]. Physical models will be tested under dynamic loading in order to explore confined pentamode lattices as effective impact mitigation devices and next-generation seismic isolation devices, with properties mainly derived from their geometric design.

**Acknowledgements**

A. Amendola gratefully acknowledges financial support from the Ph.D. School in Civil Engineering at the University of Salerno.

**REFERENCES**

[1] Milton, G.W., Cherkaev, A.V., 1995. Which elasticity tensors are realizable? J. Eng. Mater-T. 117, 4, 483-493.

[2] Hutchinson, R.G., Fleck, N.A., 2006. The structural performance of the periodic truss. J. Mech. Phys. Solids. 54 (4), 756-782.

[3] Milton, G.W., 2013. Complete characterization of the macroscopic deformations of periodic unimode metamaterials of rigid bars and pivots. J. Mech. Phys. Solids. 61 (7), 1543-1560.

[4] Norris, A.N., 2014. Mechanics of elastic networks. Proc. R. Soc. A 470, 20140522.

[5] Milton, G.W., 2013. Adaptable nonlinear bimode metamaterials using rigid bars, pivots, and actuators. J. Mech. Phys. Solids. 61 (7), 1561-1568.

[6] Martin, A., Kadic, M., Schittny, R., Bückmann, T., Wegener, M., 2010. Phonon band structures of three-dimensional pentamode metamaterials. Phys. Rev. B. 86, 155116.

[7] Huang, Y., Lu, X., Liang, G., Xu, Z., 2016. Pentamodal property and acoustic band gaps of pentamode metamaterials with different cross-section shapes. Phys. Lett. A. 380(13), 1334-1338.

[8] Bückmann, T., Thiel, M., Kadic, M., Schittny, R., Wegener, M., 2014. An elastomechanical unfeelability cloak made of pentamode metamaterials. Nat. Comm. 5,4130.

[9] Chen, Y., Liu, X., Hu, G., 2015. Latticed pentamode acoustic cloak. Scientific Reports 5:15745.

[10] Amendola, A., Smith, C.J., Goodall, R., Auricchio, F., Feo, L., Benzoni, G., Fraternali, F., 2016. experimental response of additively manufactured metallic pentamode materials confined between stiffening plates. Compos. Struct. 142, 254–262.

[11] Amendola, A., Carpentieri, G., Feo, L., Fraternali, F., 2016. Bending dominated response of layered mechanical metamaterials alternating pentamode lattices and confinement plates. Compos. Struct. 151,71-77.

[12] Fraternali, F., Carpentieri, G., Montuori, R., Amendola, A., Benzoni, G., 2015. On the use of mechanical metamaterials for innovative seismic isolation systems. In: Compdyn 2015 - 5th Eccomas thematic conference on computational methods. In: Structural Dynamics and Earthquake Engineering, pp. 349-358.

[13] Fabbrocino, F., Amendola, A., Benzoni, G., Fraternali, F., 2016. Seismic application of pentamode lattices. Ingegneria Sismica/International Journal of Earthquake Engineering, 1-2, 62-71.

[14] G. Gurtner, M. Durand, Stiffest elastic networks, Proc. R. Soc. A 470, 20130611, 2014.

[15] J. Chilton, Space Grid Structures, Oxford, UK, 2000.

[16] Kadic, M., Bückmann, T., Schittny, R., Wegener, M., 2013. On anisotropic versions of three-dimensional pentamode metamaterials. New J. Phys. 15, 023029.

[17] Destrade, M., Ogden, R.W., 2013. On stress-dependent elastic moduli and wave speeds. Ima J. Appl. Math. 78, 965-997.

[18] Meza, L.R., Das, S., Greer, J.R., 2014. Strong, lightweight and Recoverable three-dimensional ceramic nanolattices. Science. 345, 1322-1326.

[19] Zheng, X., Lee, H., Weisgraber, T.H., Shusteff, M., Deotte, J., Duoss, E.B., Kuntz, J.D., Biener, M.M., Ge, Q., Jackson, J.A., Kucheyev, S.O., Fang, N.X., Spadaccini, C.M., 2014. Ultralight, ultrastiff mechanical metamaterials. Science. 344, 61.

[20] Skinner, R.I., Robinson, W.H., Mcverry, G.H., 1993. An introduction to seismic isolation. Wiley.

[21] Kelly, J.M., 1993. Earthquake-resistant design with rubber. London: Springer-Verlag.

[22] Benzoni, G., Casarotti, C., 2009. Effects of Vertical Load, strain rate and cycling on the response of lead-rubber seismic isolators. J. Earthquake Eng. 13(3), 293-312.

[23] Higashino, M., Hamaguchi, H., Minewaki, S., Aizawa, S., 2003. Basic characteristics and durability of low-friction sliding bearings for base isolation. Earthquake Eng. Eng. Seismolog., 4(1), 95-105.

[24] EN 10210-2:2006, Hot finished structural hollow sections of non-alloy and fine grain steels - Part 2: Tolerances, dimensions and sectional properties.

[25] CSI, Computers & Structures, Inc. Analysis reference manual for SAP2000® version 18. Berkeley, California, USA, June 2015.

[26] Nesterenko, V. F., 2001. Dynamics of Heterogeneous Materials, Springer-Verlag, New York, (Chaphter 1).

[27] Fraternali, F., Senatore, L, Daraio, C., 2012, Solitary waves on tensegrity lattices. J. Mech. Phys. Solids 60, 1137-1144.

[28] Leonard, A., Ponson, L., Daraio, C., 2014. Wave mitigation in ordered networks of granular chains. J. Mech. Phys. Solids. 73, 103–117.

[29] Fraternali, F., Carpentieri, G., Amendola, A., 2015. On the mechanical modeling of the extreme softening/stiffening response of axially loaded tensegrity prisms. J. Mech. Phys. Solids. 74, 136–157.

[30] Silling, S.A., 2016, Solitary waves in a peridynamic elastic solid. J. Mech. Phys. Solids. 96, 121–132.

[31] Wang, E., Manjunath, M., Awasthi, A.P., Pal, R.K., Geubelle, P.H., Lambros, J., 2014. High-amplitude elastic solitary wave propagation in 1-D granular chains with preconditioned beads: Experiments and theoretical analysis. J. Mech. Phys. Solids. 72, 161-173.

[32] Fraternali, F., Carpentieri, G., Amendola, A. Skelton, R.E., Nesterenko, V.F., 2014. Multiscale tunability of solitary wave dynamics in tensegrity metamaterials. Appl. Phys. Lett. 105, 201903.

[33] Meaney, D.F., Morrison, B., Bass, C.D., 2014. The mechanics of traumatic brain injury: A review of what we know and what we need to know for reducing its societal burden. J. Biomech. Eng. 136(2),0210081–02100814.

[26] Nesterenko, V. F., 2001. Dynamics of Heterogeneous Materials, Springer-Verlag, New York, (Chaphter 1).

[27] Fraternali, F., Senatore, L, Daraio, C., 2012, Solitary waves on tensegrity lattices. J. Mech. Phys. Solids 60, 1137-1144.

[28] Leonard, A., Ponson, L., Daraio, C., 2014. Wave mitigation in ordered networks of granular chains. J. Mech. Phys. Solids. 73, 103–117.

[29] Fraternali, F., Carpentieri, G., Amendola, A., 2015. On the mechanical modeling of the extreme softening/stiffening response of axially loaded tensegrity prisms. J. Mech. Phys. Solids. 74, 136–157.

[30] Silling, S.A., 2016, Solitary waves in a peridynamic elastic solid. J. Mech. Phys. Solids. 96, 121–132.

[31] Wang, E., Manjunath, M., Awasthi, A.P., Pal, R.K., Geubelle, P.H., Lambros, J., 2014. High-amplitude elastic solitary wave propagation in 1-D granular chains with preconditioned beads: Experiments and theoretical analysis. J. Mech. Phys. Solids. 72, 161-173.

[32] Fraternali, F., Carpentieri, G., Amendola, A. Skelton, R.E., Nesterenko, V.F., 2014. Multiscale tunability of solitary wave dynamics in tensegrity metamaterials. Appl. Phys. Lett. 105, 201903.

[33] Meaney, D.F., Morrison, B., Bass, C.D., 2014. The mechanics of traumatic brain injury: A review of what we know and what we need to know for reducing its societal burden. J. Biomech. Eng. 136(2),0210081–02100814.